\documentclass[10pt]{iopart}

\usepackage{iopams}  
\usepackage{graphicx}
\usepackage[breaklinks=true,colorlinks=true,linkcolor=blue,urlcolor=blue,citecolor=blue]{hyperref}

\begin{document}

\title[Uniaxial anisotropy and two-magnon scattering in Co$_{2}$FeSi] {Uniaxial anisotropy, intrinsic and extrinsic damping in Co$_{2}$FeSi Heusler alloy thin films}

\author{Binoy Krishna Hazra$^{1}$, S. N. Kaul$^{1}$, S. Srinath$^{1}$}
\address{$^1$School of Physics, University of Hyderabad, Hyderabad-500046, India}

\author{M. Manivel Raja$^{2}$}
\address{$^{2}$Defence Metallurgical Research Laboratory, Hyderabad-500058, India}

\eads{\mailto{sn.kaul@uohyd.ac.in} and \mailto{srinath@uohyd.ac.in}}

\begin{abstract}

Ferromagnetic resonance (FMR) technique has been used to study the magnetization relaxation processes and magnetic anisotropy in two different series of the Co$_{2}$FeSi (CFS) Heusler alloy thin films, deposited on the Si(111) substrate by ultra high vacuum magnetron sputtering. While the CFS films of fixed (50 nm) thickness, deposited at different substrate temperatures ($T_{S}$) ranging from room temperature (RT) to 600$^\circ$C, constitute the series-I, the CFS films with thickness $t$ varying from 12 nm to 100 nm and deposited at 550$^\circ$C make up the series-II. In series-I, the CFS films deposited at T$_{S}$ = RT and 200$^\circ$C are completely amorphous, the one at T$_{S}$ = 300$^\circ$C is partially crystalline, and those at $T_{S}$ = 450$^\circ$C, 550$^\circ$C and 600$^\circ$C are completely crystalline with B2 order. By contrast, all the CFS films in series-II are in the fully-developed B2 crystalline state. Irrespective of the strength of disorder and film thickness, angular variation of the resonance field in the film plane unambiguously establishes the presence of \textit{global} `in-plane' uniaxial anisotropy. The uniaxial anisotropy field decreases as the crystalline order in the films increases and goes through a minimum at $t$ = 50 nm as a function of film thickness. Landau$-$Lifshitz$-$Gilbert damping and two-magnon scattering dominantly contribute to the line-broadening in both `in-plane' and `out-of-plane' configurations. The two-magnon scattering has larger magnitude in the amorphous films than in the crystalline ones. Angular variation of the linewidth in the film plane reveals that, in the CFS thin films of varying thickness, a crossover from the `in-plane' \textit{local} four-fold symmetry (cubic anisotropy) to \textit{local} two-fold symmetry (uniaxial anisotropy) occurs as $t$ exceeds 50 nm. Gilbert damping parameter $\alpha$ decreases monotonously from 0.047 to 0.0078 with decreasing disorder strength (increasing T$_{S}$) and jumps from 0.008 for the CFS film with $t$ = 50 nm to 0.024 for the film with $t$ = 75 nm. Such variations of $\alpha$ with T$_{S}$ and $t$ are understood in terms of the changes in the total (spin-up and spin-down) density of states at the Fermi level caused by the disorder and film thickness. Spin pumping across the Co$_{2}$FeSi film/Ta cap-layer interface makes negligible contribution to $\alpha$. We propose that disorder and/or the film thickness can be used as control parameters to tune $\alpha$, whose value is decisive in choosing a ferromagnetic film for a given spintronics application.

\end{abstract}



\section{INTRODUCTION}



Recognition of the fundamental role of spin polarization and magnetization dynamics in spintronics devices has led to a surge in the number of investigations of the magnetic relaxation phenomenon in thin films with long-range magnetic order. Among various experimental techniques, ferromagnetic resonance (FMR) has been widely used to investigate the relaxation processes in magnetic thin films. Magnetic anisotropy constant, Land\'{e} g-factor, Gilbert damping parameter are extracted from the resonance field and FMR linewidth. The Gilbert damping parameter ($\alpha$) is, by far, the most crucial factor in deciding whether or not a given magnetic system is an appropriate choice for spintronics devices \cite{TanjaSSC2011}. For instance, a small value of $\alpha$ is required for magnetic tunnel junction devices and spin-torque transfer-based devices; in the latter case, for reducing the current density (and hence minimizing the power consumption) required for the spin transfer torque switching \cite{BergerPRB1996, SlonczewskiJMMM1996}. On the other hand, a large value of $\alpha$ is needed to improve thermal stability in current-perpendicular-to-plane giant magnetoresistance (GMR) read sensors \cite{SmithAPL2001, LiuJAP2011}. The magnitude of spin current that can be generated using spin pumping, is inversely proportional to the Gilbert damping parameter \cite{AndoNature2011, KurebayashiAPL2013}. Furthermore, Gilbert damping is directly related to density of the states at the Fermi level \cite{Kambersky1970, KubotaAPL2009, MizukamiJAP2009, OoganeAPL2010, SakumaJOPD2015, Schoen2016}. Though there are indications that the anti-site disorder substantially changes $\alpha$ in many Heusler alloy thin films, systematic investigations of how disorder affects $\alpha$ are still lacking. A low value of the Gilbert damping parameter $\alpha$ = 0.0022 was reported \cite{YangAPL2013} for the L2$_{1}$ ordered Co$_{2}$FeSi film after post-deposition annealing at 650$^{\circ}$C. In sharp contrast, a recent investigation of $\alpha$ in Co-based Heusler alloys reveals that as-deposited amorphous Co$_{2}$FeSi films and the B2 ordered films annealed at 300$^{\circ}$C have a lower $\alpha$ (0.008) compared to the relatively large $\alpha$ (0.04) of the L2$_{1}$ ordered film annealed at 400$^{\circ}$C \cite{OoganeJOPD2015}. In the same study, it was shown that, as a function of annealing temperature, $\alpha$ goes through a minimum at 300$^{\circ}$C in Co$_{2}$MnAl and Co$_{2}$MnSi films. 

The Gilbert damping parameter $\alpha$ for Co-based Heusler alloys has been determined either from the angular variation of the FMR linewidth at a fixed frequency or from the frequency dependence of linewidth or both \cite{MizukamiJAP2009, Schoen2016, BelmeguenaiPRB2013, KoehlerPRB2016, SterwerfCondmat2016, BainslaJOPD2018}. The measured FMR linewidth has contributions from the \textit{intrinsic} (Gilbert damping) as well as \textit{extrinsic} (such as the defect-induced two-magnon scattering, inhomogeneity in magnetization, angular spread of crystallite misorientation and spin pumping) linewidth broadening mechanisms. Thus, only when the relative magnitudes of all the contributions to FMR linewidth are unambiguously determined, FMR linewidth can provide accurate determination of $\alpha$ and a quantitative measure of the degree of magnetic inhomogeneity \cite{HeinrichPRL1987, PlatowPRB1998} in a given system. By not considering one or more of the extrinsic linewidth broadening contributions, many of the earlier investigations of magnetic relaxation process in Cobalt-based Heusler alloys \cite{ MizukamiJAP2009, YangAPL2013, OoganeJOPD2015, BelmeguenaiPRB2013, KoehlerPRB2016, SterwerfCondmat2016, BainslaJOPD2018} are somewhat flawed. The conflicting results \cite{YangAPL2013, OoganeJOPD2015} about the effect of crystal structure on $\alpha$ may also be a consequence of a total neglect of some of the extrinsic linewidth contributions.    

In this work, the effect of disorder and film thickness on magnetic anisotropy and Gilbert damping in Co$_{2}$FeSi Heusler alloy thin films is brought out unambiguously from the observed angular variations of the resonance field and linewidth in the `in-plane' and `out-of-plane' sample configurations. In addition to the intrinsic Landau-Lifshitz-Gilbert (LLG) damping, all possible extrinsic magnetic relaxation mechanisms (stated above) are considered in order to accurately determine $\alpha$.

\section{EXPERIMENTAL DETAILS}

Ultra high vacuum dc magnetron sputtering was used to deposit two different series of Co$_{2}$FeSi (CFS) Heusler alloy thin films on the Si (111) substrate. The series I comprises CFS thin films of fixed thickness 50 nm, deposited at the constant substrate temperatures T$_{S}$ = 27 $^{\circ}$C (room temperature, RT), 200 $^{\circ}$C, 300 $^{\circ}$C, 450 $^{\circ}$C, 550 $^{\circ}$C and 600 $^{\circ}$C and labeled as the RT, TS200, TS300, TS450, TS550 and TS600 films. The CFS thin films of thicknesses 12 nm, 25 nm, 50 nm, 75 nm and 100 nm, deposited at the fixed substrate temperature T$_{S}$ = 550 $^{\circ}$C, constitute the series II. The details of the film deposition conditions and parameters can be found elsewhere \cite{BKH2017}. Previously reported \cite{BKH2017,BKH2019} grazing-incidence x-ray diffraction (GIXRD) patterns, taken on the films of fixed thickness (50 nm), have revealed the amorphous nature of the CFS films deposited at room temperature (RT) and 200$^{\circ}$C and the initiation of crystallinity in the otherwise amorphous film deposited at 300$^{\circ}$C. The remaining films, deposited at 450$^{\circ}$C, 550$^{\circ}$C and 600$^{\circ}$C, are in the fully-developed crystalline state but the one deposited at 550$^{\circ}$C has the most well-defined B2 ordered structure. Thus, the CFS films of different thicknesses (series II) were deposited at T$_{S}$ = 550$^{\circ}$C. While the CFS thin films with varying degree of disorder (i.e., the RT, TS200, TS300, TS450, TS550 and TS600 thin films; series I) permit an in-depth study of the effect of disorder, the series II enables determination of how the film thickness affects the magnetic anisotropy, Gilbert damping and the extrinsic line-broadening contributions. Another important aspect of this work is that the deposition at fixed substrate temperatures is preferred to the customary practice of annealing the films at different temperatures after they have been deposited at room temperature. The rationale behind this preference is that the post-deposition annealing promotes inter-layer diffusion between the film, the cap-layer and/or the buffer layer, with the result the investigations involving post-deposition annealing often lead to inconclusive results. On the contrary, in the present case, only after the substrate (and hence the deposited film) had cooled down to room temperature in the ultra high vacuum chamber, the CFS film was capped with 2 nm Ta layer \cite{BKH2017}. This arrangement not only prevents surface oxidation but also the film/cap-layer inter-diffusion. 

To investigate the type of magnetic anisotropy present in Co$_2$FeSi (CFS) films, and to unambiguously separate out the intrinsic and extrinsic contributions to the FMR linewidth and accurately determine their relative magnitudes, FMR spectra have been recorded at various azimuthal (`in-plane') field angles, $\varphi_{H}$, and polar (`out-of-plane') field angles, $\theta_{H}$, on the square-shaped CFS thin films. The `in-plane' and `out-of-plane' sample-mounting configurations as well as the polar coordinate system, defining the `in-plane' (`out-of-plane') field and magnetization angles, $\varphi_{H}$ and $\varphi_{M}$ ($\theta_{H}$ and $\theta_{M}$), are shown in figure 1.

\begin{figure*}[htbp]
\centering
\includegraphics[scale=1.0, trim = 0 0 0 0, clip, width=0.5\linewidth]{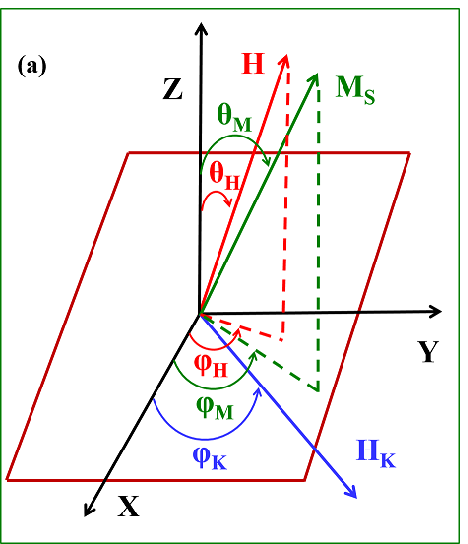}\\
\hspace{0.5cm}
\includegraphics[scale=1.0, trim = 0 0 0 0, clip, width=0.40\linewidth]{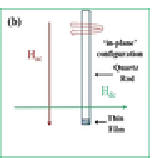}
\hspace{0.5cm}
\includegraphics[scale=1.0, trim = 0 0 0 0, clip, width=0.405\linewidth]{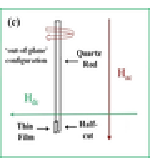}
\caption{(a): Schematic representation of the coordinate system pertinent to the ‘in-plane’ (IP) and ‘out-of-plane’
(OP) FMR measurements. (b) and (c): Thin films mounted on solid quartz rod and half-cut quartz rod in IP and OP configurations. The directions of the alternating ($H_{ac}$) and static ($H_{dc}$) magnetic fields are also indicated.}
\label{figure 1}
\vspace{-0.5cm}
\end{figure*}

\section{THEORETICAL CONSIDERATIONS}

\subsection{Angular variation of resonance field}

In the coordinate system, shown in figure 1, $xy$ plane represents the film plane and $z$ axis is normal to the film plane. The angle between the magnetic field, $H$, (magnetization, $M$) direction and the film normal is denoted by $\theta_{H}$ ($\theta_{M}$). The projection of the magnetic field (magnetization) on $xy$ plane makes an angle $\varphi_{H}$ ($\varphi_{M}$) with the $x$ axis. The anisotropy field is assumed to be in the $xy$ plane, making an angle $\varphi_{K}$ with $x$ axis.

Therefore, the total energy can be written as \cite{Siruguri1996}
\begin{eqnarray}
\fl
F=-MH~[sin~\theta_{M}~sin~\theta_{H}~cos~(\varphi_{M}-\varphi_{H})+cos~\theta_{M}~cos~\theta_{H}] + 2\pi~M^{2}~cos^{2}~\theta_{M} \nonumber \\
+ K_{u}~[1-sin^{2}~\theta_{M}~cos^{2}~(\varphi_{M}-\varphi_{K})]
\label{Eq 1}
\end{eqnarray}

The equilibrium conditions for magnetization can be found from $\partial F/\partial \theta_{M}=0$ and $\partial F/\partial \varphi_{M}=0$. The resonance condition, which relates the free energy with the resonance frequency, is \cite{SuhlPR1955}

\begin{eqnarray}
\fl
\left(\frac{\omega}{\gamma}\right)^{2}=\frac{1}{M^{2}~sin^{2}~\theta_{M}}\left[ \left( \frac{\partial^2 F}{\partial \theta_{M}^2}~\frac{\partial^2 F}{\partial \varphi_{M}^2}\right) -\left( \frac{\partial^2 F}{\partial \theta_{M}\partial \varphi_{M}} \right)^{2}\right] 
\label{Eq 2}
\end{eqnarray}

In the `in-plane' (IP) case, both $M$ and $H$ vectors are confined to the sample plane ($xy$ plane), i.e., $\theta_{M}=\theta_{H} =\frac{\pi}{2}$ and the angles $\varphi_{M}$, $\varphi_{H}$ and $\varphi_{k}$ are measured with respect to the $x$-axis in the sample plane. The resonance condition is given by \cite{Siruguri1996, BasheedJAP2011}

\begin{eqnarray}
\fl
\left(\frac{\omega}{\gamma}\right)^{2}= H^{\parallel}_{1} \times H^{\parallel}_{2}
\label{Eq 3}
\end{eqnarray}

where $H^{\parallel}_{1}$ = $[H_{r}^{\parallel}~cos~(\varphi_{H}-\varphi_{M})+(4\pi M+H_{k}^{\parallel}~cos^{2}~\varphi_{M})]$ and $H^{\parallel}_{2}$ = $[H_{r}^{\parallel}~cos~(\varphi_{H}-\varphi_{M})+H_{k}^{\parallel}~cos~2\varphi_{M})]$ denote the \textit{stiffness fields} for the IP case.

Equilibrium condition for magnetization [$\partial F/\partial \varphi_{M}=0$] has the form \cite{Siruguri1996, BasheedJAP2011}

\begin{eqnarray}
\fl
2 H_{r}^{\parallel}~sin~(\phi_{H}-\phi_{M})=H_{k}^{\parallel}~sin~(2 \phi_{M})
\label{Eq 4}
\end{eqnarray}

In the `out-of-plane' (OP) case, $H$, $M$ and $H_{K}$ are confined to the $xz$ plane so that $\varphi_{H}=\varphi_{M}=\varphi_{K} = 0$ and the angles $\theta_{H}$ and $\theta_{M}$ are measured with respect to the $z$-axis (normal to the film plane). The OP resonance condition is \cite{Siruguri1996, BasheedJAP2011} 

\begin{eqnarray}
\fl
(\omega/\gamma)^{2} = H^{\perp}_{1} \times H^{\perp}_{2} 
\label{Eq 5}
\end{eqnarray}

where, $H^{\perp}_{1}$ = $[H_{r}^{\perp}~cos~(\theta_{H}-\theta_{M})+(4\pi M+H_{k}^{\perp})~cos~2\theta_{M}]$ and $H^{\perp}_{2}$ = $[H_{r}^{\perp}~cos~(\theta_{H}-\theta_{M})-4\pi M~cos^{2}~\theta_{M}+H_{k}~sin^{2}~\theta_{M}]$ represent the \textit{stiffness fields} in the OP sample configuration.

The corresponding equilibrium condition for magnetization [$\partial F/\partial \theta_{M}=0$] is \cite{Siruguri1996, BasheedJAP2011}

\begin{eqnarray}
\fl
\frac{cos~\theta_{H}}{cos~\theta_{M}}-\frac{sin~\theta_{H}}{sin~\theta_{M}} = \frac{4\pi M_{s}+H_{K}^{\perp}}{H_{r}^{\perp}}
\label{Eq 6}
\end{eqnarray}

with $4 \pi M_{eff}$ = ${4 \pi M_{s} + H_{K}^{\perp}}$, where $M_{s}$ is the saturation magnetization.

\subsection{Angular variation of linewidth}

The `peak-to-peak' ferromagnetic resonance linewidth, $\Delta H$, is a measure of the rate at which magnetization relaxes back to equilibrium once the static magnetic field is switched off. In the Arias-Mills approach \cite{AriasPRB1999, LanderosPRB2008, LinderPRB2009, BarsukovPRB2012}, which considers both the intrinsic and extrinsic damping mechanisms, FMR linewidth can be written as 

\begin{eqnarray}
\fl
\Delta H = \Delta H^{LLG}+\Delta H^{TMS}
\label{Eq 7}
\end{eqnarray}

The first term in the expression for $\Delta H$ (Eq.(7)) denotes the \textit{intrinsic} Gilbert damping contribution, which is proportional to the microwave field frequency ($\omega$)

\begin{eqnarray}
\fl
\Delta H^{LLG}= \frac{2}{\sqrt{3}}~ \frac{\alpha ~\omega}{\gamma ~M_{s}~ \Xi}
\label{Eq 8}
\end{eqnarray}

In Eq.(8), the prefactor $\frac{2}{\sqrt{3}}$ is needed to obtain the `peak-to-peak' linewidth and $\alpha$ is the Gilbert damping parameter. A deviation of external field angle from the equilibrium magnetization angle leads to the dragging of magnetization by field and is expressed in terms of the \textit{dragging function}, defined as

\begin{eqnarray}
\fl
\Xi = \frac{1}{(H_{1}+H_{2})} \frac{d(\omega/\gamma)^{2}}{dH_{r}}
\label{Eq 9}
\end{eqnarray}

where H$_{1}$ and H$_{2}$ are the \textit{stiffness fields}, defined earlier. The second term ($\Delta H^{TMS}$) in Eq.(7) represents the \textit{extrinsic} damping contribution, arising from the two-magnon scattering (TMS). $\Delta H^{TMS}$ is given by the following expression

\begin{eqnarray}
\fl
\Delta H^{TMS}= \frac{\Gamma}{\gamma ~\Xi}
\label{Eq 10}
\end{eqnarray}

$\Gamma$ is the two-magnon scattering rate, which depends on the microwave field frequency and the static magnetic field angle. The TMS originates from the nonuniform magnon modes with wavevector $\textbf{q} \neq 0$ \cite{HurbenJAP1998, HeinrichPRB2004}.

Besides the Gilbert and TMS contributions, the inhomogeneity in magnetization and the angular spread of crystallite misorientation cause additional broadening of the linewidth \cite{ChappertPRB1986, MizukamiJJAP2001}. These contributions to $\Delta H$ are given by 

\begin{eqnarray}
\fl
\Delta H^{4 \pi M_{eff}}=\left\vert \frac{\partial H_{r}}{\partial (4 \pi M_{eff})} \right\vert ~\Delta (4 \pi M_{eff})
\label{Eq 11}
\end{eqnarray} 
and 
\begin{eqnarray}
\fl
\Delta H^{\xi_{H}}=\left\vert \frac{\partial H_{r}}{\partial \xi_{H}} \right\vert ~\Delta (\xi_{H})
\label{Eq 12}
\end{eqnarray}

where $\xi_{H}$ stands for $\varphi_{H}$ or $\theta_{H}$, while $\Delta (4 \pi M_{eff})$ and $\Delta (\xi_{H})$ are the average spread in $4 \pi M_{eff}$, and in the easy (preferred) direction of magnetization within the film plane, dictated by the magnetic anisotropy. 

Combining Eqs.(3) or (5), (4) or (6), (8) - (12), yields the final expressions for different contributions to FMR linewidth as \cite{BelmeguenaiPRB2013, MillsPB2006, AriasJAP2000}

\begin{eqnarray}
\fl
\Delta H^{LLG}= \frac{2}{\sqrt{3}} ~\frac{\alpha~ \omega}{\gamma~ M_{s}}~~...~~for ~~(IP)\\
\fl
\Delta H^{LLG}= \frac{2}{\sqrt{3}}~\frac{\alpha ~\omega}{\gamma ~M_{s}~cos~(\theta_{H}-\theta_{M})}~~...~~for ~~(OP)
\label{Eq 14}
\end{eqnarray}

\begin{eqnarray}
\fl
\Delta H^{TMS}=[\Gamma_{0}+\Gamma_{2}~cos~2(\varphi_{H}-\varphi_{2})+\Gamma_{4}~cos~4(\varphi_{H}-\varphi_{4})]\nonumber\\ ~~~~~~~~~~~~~~~~~~~~\times arcsin\left(\frac{f}{\sqrt{f^{2}+f_{0}^{2}}+f_{0}} \right) ~~...~~for ~~(IP)\\
\fl 
\Delta H^{TMS} = \frac{2}{\sqrt{3}} ~\Gamma (H_{0},\theta_{H}) ~ sin^{-1} \sqrt{\frac{H_{1}^{\perp}}{H_{1}^{\perp}+M_{eff}} \frac{cos~(2\theta_{M})}{cos^{2}~\theta_{m}}} ~~...~~for ~~(OP)
\label{Eq 16}
\end{eqnarray}

where $f_{0}=\gamma~ M_{eff}$, $\Gamma_{2}$ and $\Gamma_{4}$ are the strengths of the magnetic anisotropies of the two-fold and four-fold symmetry, respectively.

\begin{eqnarray}
\fl
\Delta H^{4 \pi M_{eff}} = \frac{H_{2}^{\parallel}}{(H_{1}^{\parallel}+H_{2}^{\parallel})~cos~(\varphi_{H}-\varphi_{M})}~ \Delta (4 \pi M_{eff}) ~~...~~for ~~(IP)\\
\fl
\Delta H^{4 \pi M_{eff}} = \frac{H_{2}^{\perp}~sin^{2}~\theta_{m}-H_{1}^{\perp}~cos~2\theta_{m}}{(H_{1}^{\perp}+H_{2}^{\perp})~cos~(\theta_{H}-\theta_{M})}~\Delta (4 \pi M_{eff}) ~~...~~for ~~(OP)
\label{Eq 18}
\end{eqnarray}  

\begin{eqnarray}
\fl
\Delta H^{\varphi_{H}} = H_{r}^{\parallel}~tan~(\varphi_{H}-\varphi_{M}) ~\Delta \varphi_{H} ~~...~~for ~~(IP) \\
\fl
\Delta H^{\theta_{H}} = H_{r}^{\perp}~tan~(\theta_{H}-\theta_{M})~\Delta \theta_{H}~~...~~for ~~(OP)
\label{Eq 20}
\end{eqnarray}

\begin{figure*}[htbp]
\centering
\includegraphics[scale=1.0, trim = 0 70 0 100, clip, width=\linewidth]{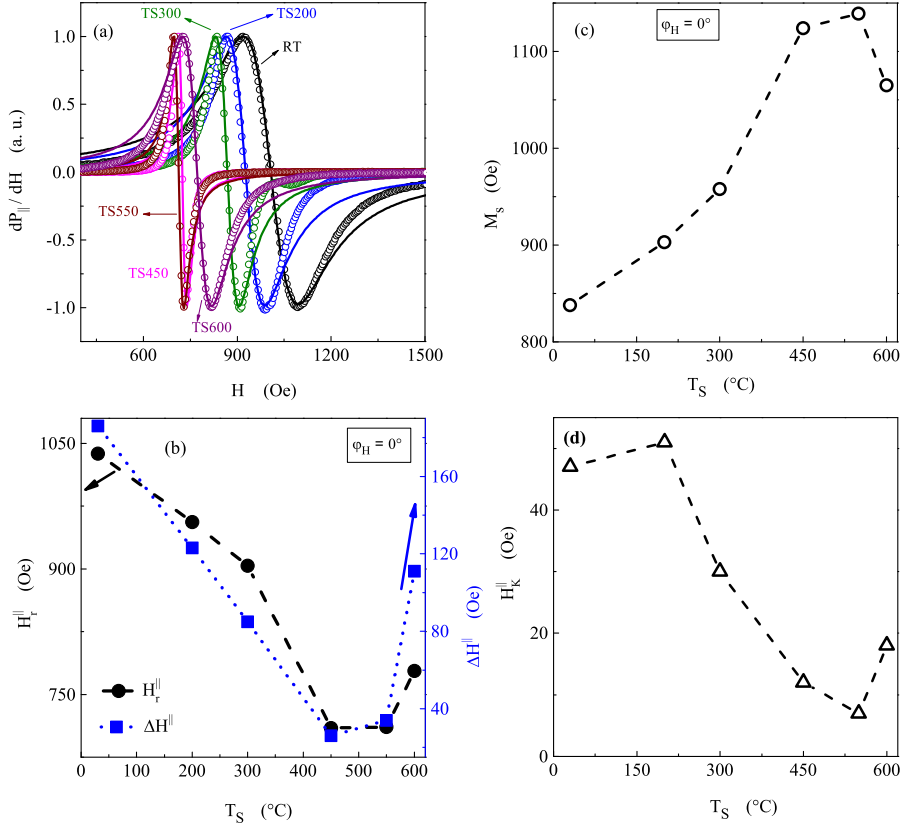}
\caption{ (a): Static magnetic field dependence of the field derivative of the resonant microwave power absorption in the IP sample configuration, $\frac{dP_{\parallel}}{dH}$, at a fixed field angle ($\varphi_{H} = 0^{\circ}$) in the 50 nm CFS films, deposited at different T$_{S}$. (b) - (d): Variations of the resonance field, H$_{r}^{\parallel}$, FMR linewidth, $\Delta$H$\parallel$, saturation magnetization, M$_{s}$ and 'in-plane' anisotropy field, H$_{K}^{\parallel}$, with T$_{S}$ at the field angle $\varphi_{H} = 0^{\circ}$ in the parallel sample geometry for the CFS films of fixed thickness (50 nm).}
\label{figure 2}
\vspace{-0.5cm}
\end{figure*}

\begin{figure*}[htbp]
\centering
\includegraphics[scale=1.0, trim = 0 200 0 200, clip, width=\linewidth]{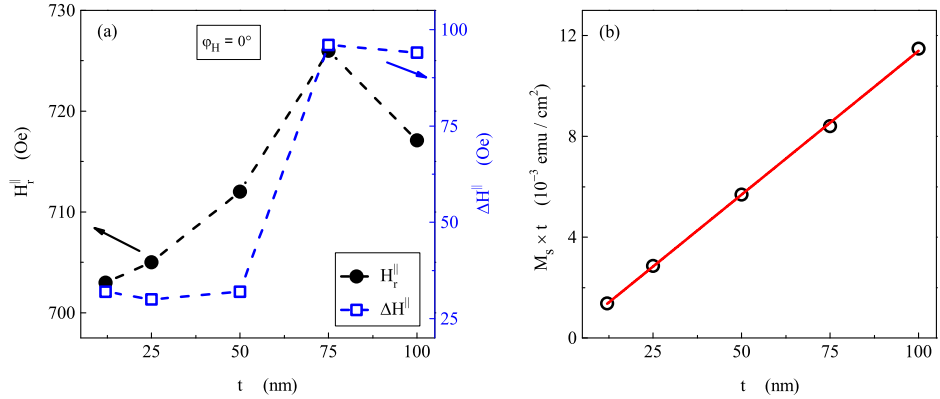}
\caption{Variation of (a) the resonance field and FMR linewidth, and (b) M$_{s}$ $\times$ $t$, with the film thickness, $t$, at fixed field angle $\varphi_{H} = 0^{\circ}$ in the parallel geometry for the CFS films of different thicknesses.}
\label{figure 3}
\vspace{-0.5cm}
\end{figure*}

\section{RESULTS and DISCUSSION} 

\subsection{`In-plane' (IP) configuration}  

Derivative of the microwave power absorption with respect to static magnetic field (H), in the \textit{parallel configuration} [i.e., when the angle $\varphi_{H}$ between static magnetic field and the $x$-axis in the $xy$ (sample) plane, as shown in figure 1(a), equals zero], $\left(\frac{dP_{\parallel}}{dH}\right)$, has been recorded at room temperature on the Co$_{2}$FeSi (CFS) films belonging to the series I and II. The $\frac{dP}{dH}$ versus H curves, recorded at $\varphi_{H}$ = 0$^{\circ}$ in the `in-plane' configuration on the thin films comprising the series I, representative of the series II as well, are depicted in figure 2(a) along with the theoretical fits (continuous curves) yielded by the lineshape (LS) analysis \cite{KaulJPF1987,KaulJPCM1992,BabuPRB1992,SiruguriJPCM1996}, based on the Landau-Lifshitz-Gilbert (LLG) equation of motion for dynamic magnetization. To arrive at these optimum fits, saturation magnetization (M$_{s}$), Land\'{e} splitting factor ($g$), anisotropy field (H$_{k}^{\parallel}$) and FMR linewidth ($\Delta H^{\parallel}$) are treated as free-fitting parameters. Regardless the value of T$_{S}$ (or the degree of disorder present) and/or film thickness, the Land\'{e} splitting factor has the value $g$ = 2.04(2).

The resonance field, H$_{r}^{\parallel}$, $\Delta H^{\parallel}$, M$_{s}$ and H$_{k}^{\parallel}$, obtained from the LS analysis, are plotted as functions of T$_{S}$ in Fig.2(b) - (c). While H$_{r}^{\parallel}$, $\Delta H^{\parallel}$, and H$_{k}^{\parallel}$ go through a \textit{minimum} at T$_{S}$ = 550$^{\circ}$C, M$_{s}$ \textit{peaks} at the same T$_{S}$ value. The numerical values of M$_{s}$ and H$_{k}^{\parallel}$ are listed in Table 1. Considering that the FMR linewidth is a measure of the magnetic \textit{inhomogenity} in a ferromagnetic system \cite{PlatowPRB1998,FarlePRB1996}, the decrease in $\Delta H^{\parallel}$ and increase in M$_{s}$ with increasing T$_{S}$ up to 550$^{\circ}$C obviously stem from the suppression of anti-site disorder and the consequent progressive improvement in crystallographic B2 ordering \cite{BKH2019}. The subsequent increase (decrease) in $\Delta H^{\parallel}$ (M$_{s}$) at T$_{S}$ $>$ 550$^{\circ}$C, most likely, results from the Si inter-diffusion at the interface between the CFS film and the Si substrate. Similar variation of M$_{s}$ with the post-deposition annealing temperature is observed in many cobalt-based Heusler alloys \cite{OoganeJOPD2015, BelmeguenaiJAP2014}.  The opposite variations of H$_{r}^{\parallel}$ and M$_{s}$ with T$_{S}$, evident in Fig.2(b) and 2(c), can be understood in terms of the Kittel FMR condition \cite{Kittel1996} which relates H$_{r}$ and M$_{s}$ as $(\omega/\gamma)^{2}= H_{r}^{\parallel}\left( H_{r}^{\parallel}+4 \pi M_{s}\right)$.   

In amorphous ferromagnets, magnetization does not saturate even in static magnetic fields, H$_{dc}$, as large as a few hundreds of kOe because of the distribution in local magnetization prevalent in them. On the other hand, in FMR experiments, the measured value for M$_{s}$ corresponds to fields H$_{dc}$ = H$_{r}^{\parallel}$ $\simeq$ 1 kOe (Fig.3(a)). FMR, thus, invariably measures a lower value of M$_{s}$ in the amorphous films of given composition and thickness than that obtained from the bulk magnetization measurements where much stronger fields are applied in order to saturate the magnetization. Thus, it is not surprising that, in the crystalline films, where magnetization can be saturated at much lower fields, FMR and direct magnetization measurements yield nearly identical values for the saturation magnetization \cite{BeaujourJAP2008, WeiJOPD2013}. 

Figure 3 depicts the variations of H$_{r}^{\parallel}$, $\Delta H^{\parallel}$ and M$_{s}$ $\times$ $t$ with the film thickness, $t$. Fig.3(b) demonstrates that the M$_{s}$ $\times$ $t$ versus $t$ plot is linear with 0.12 nm as the intercept on the thickness axis. Note that $t_{dl}$ = 0.12 nm is the thickness of the magnetically 'dead' layer because  M$_{s}$ = 0 in this layer. That M$_{s}$ obeys the relation M$_{s} = a + b/t$ has also been previously reported \cite{BelmeguenaiPRB2013, KoehlerPRB2016, BelmeguenaiJMMM2014} in Co$_{2}$FeAl and Co$_{2-x}$Ir$_{x}$MnSi Heusler-alloy thin films. H$_{K}^{\parallel}$ goes through a minimum at $t$ = 50 nm (not shown in Fig.3; see Table 2 for the actual values of H$_{K}^{\parallel}$). 

\begin{figure*}[htbp]
\centering
\includegraphics[scale=1.0, trim = 0 0 0 0, clip, width=\linewidth]{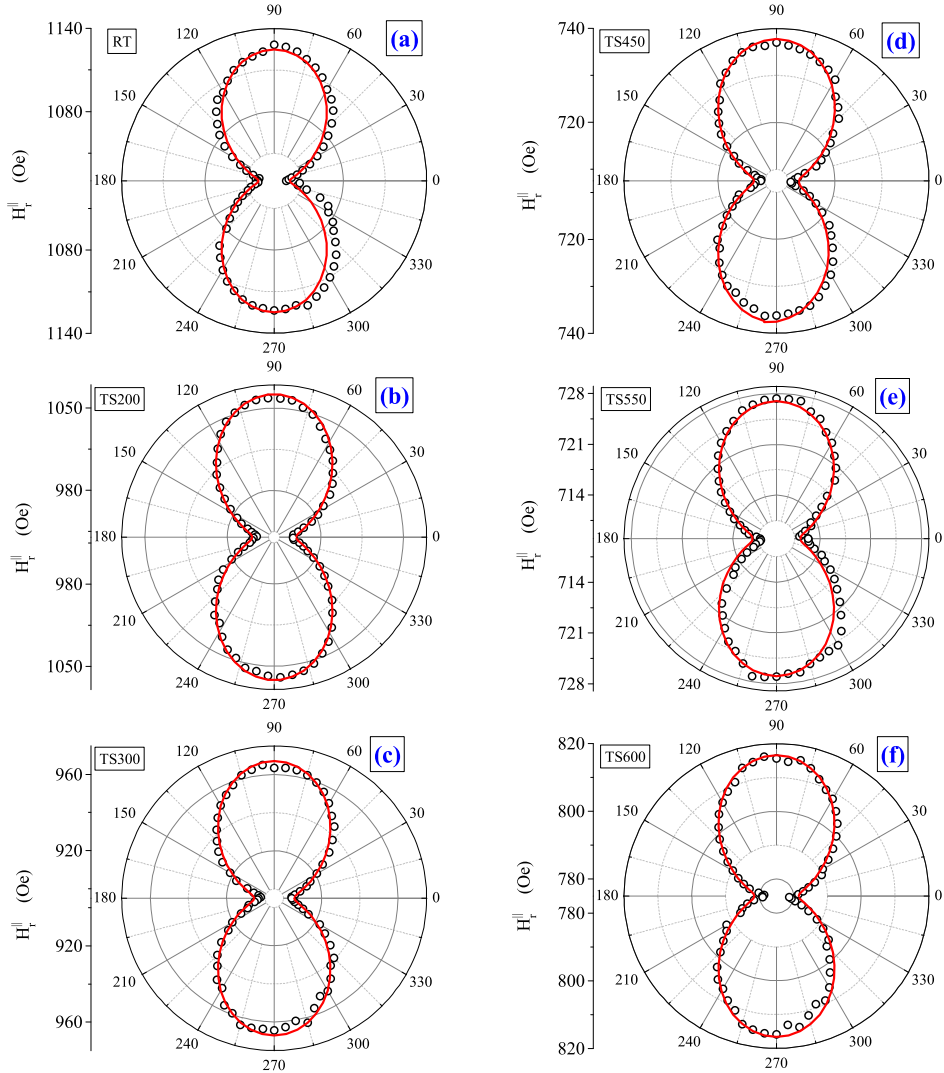}
\caption{Polar plots of the resonance field, H$_{r}^{\parallel}$, versus $\varphi_{H}$ for the 50 nm thick CFS films at T = 300 K in the IP configuration. The continuous curves through the data points represent the best least-squares fits based on Eqs.(3) and (4).}
\label{figure 4}
\vspace{-0.5cm}
\end{figure*}

\begin{figure*}[htbp]
\centering
\includegraphics[scale=1.0, trim = 0 100 0 70, clip, width=\linewidth]{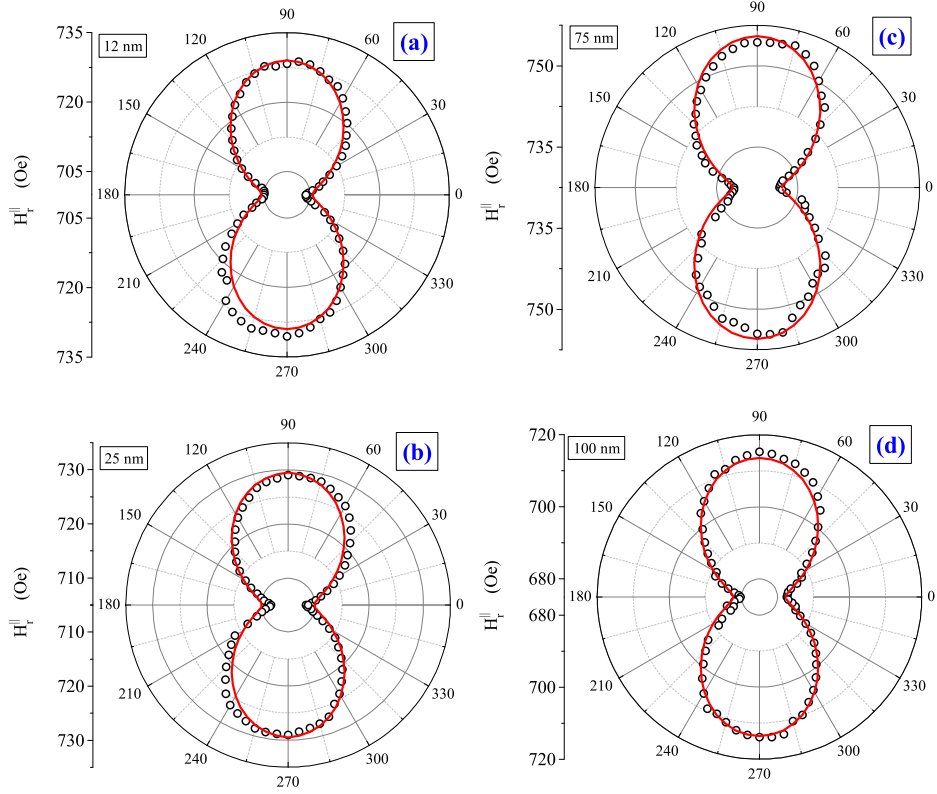}
\caption{Polar plots of resonance field, H$_{r}^{\parallel}$, versus $\varphi_{H}$ for the CFS films of different thicknesses at T = 300 K in the IP configuration. The continuous curves through the data points represent the best least-squares fits, based on Eqs.(3) and (4).}
\label{figure 5}
\vspace{-0.5cm}
\end{figure*}

\begin{figure*}[htbp]
\centering
\includegraphics[scale=1.0, trim = 0 250 20 200, clip, width=\linewidth]{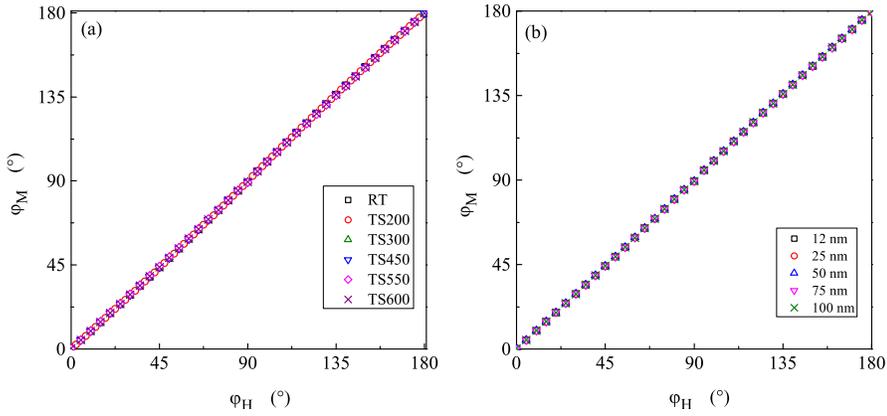}
\caption{Variation of the IP equilibrium magnetization angle, $\varphi_{M}$, with the IP field angle, $\varphi_{H}$, for (a) the 50 nm CFS thin films and (b) the CFS films with thickness in the range 12-100 nm.}
\label{figure 6}
\vspace{-0.5cm}
\end{figure*}

\begin{figure*}[htbp]
\centering
\includegraphics[scale=1.0, trim = 0 0 0 0, clip, width=\linewidth]{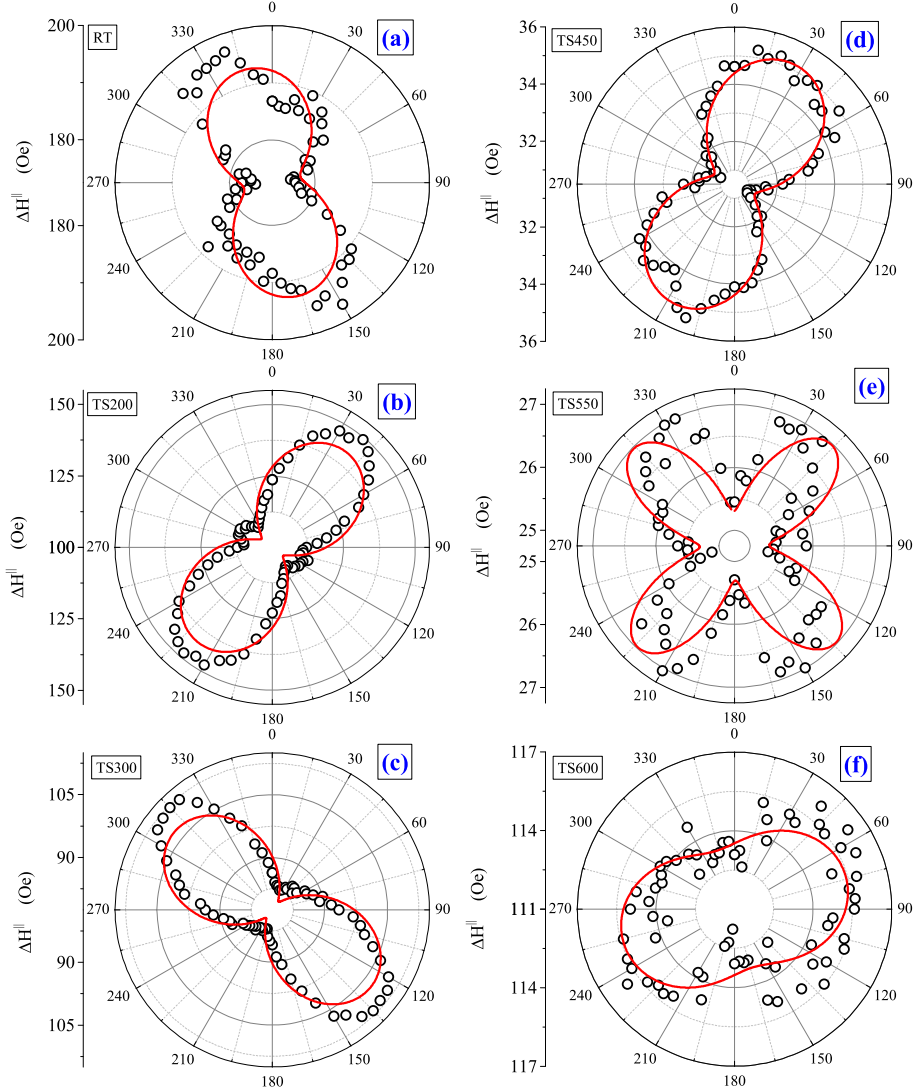}
\caption{Polar plots of the FMR linewidth, $\Delta H^{\parallel}$, versus $\varphi_{H}$ for the 50 nm thick CFS films at room temperature in the IP configuration. The continuous curves through the data points represent the theoretical fits based on Eqs.(13), (15), (17) and (19).}
\label{figure 7}
\vspace{-0.5cm}
\end{figure*}

\begin{figure*}[htbp]
\centering
\includegraphics[scale=1.0, trim = 0 100 0 50, clip, width=\linewidth]{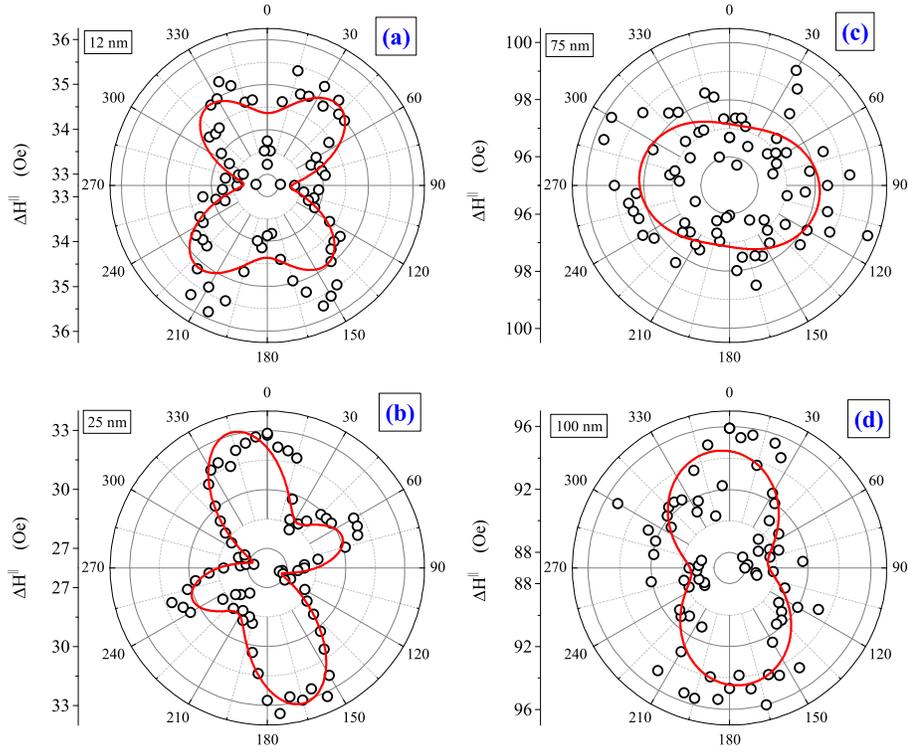}
\caption{Polar plots of FMR linewidth, $\Delta H^{\parallel}$, versus $\varphi_{H}$ for the CFS films of different thicknesses at room temperature in the IP configuration. The continuous curves through the data points represent the theoretical fits based on Eqs.(13), (15), (17) and (19).}
\label{figure 8}
\vspace{-0.5cm}
\end{figure*}

\begin{figure*}[htbp]
\centering
\includegraphics[scale=1.0, trim = 0 0 0 0, clip, width=\linewidth]{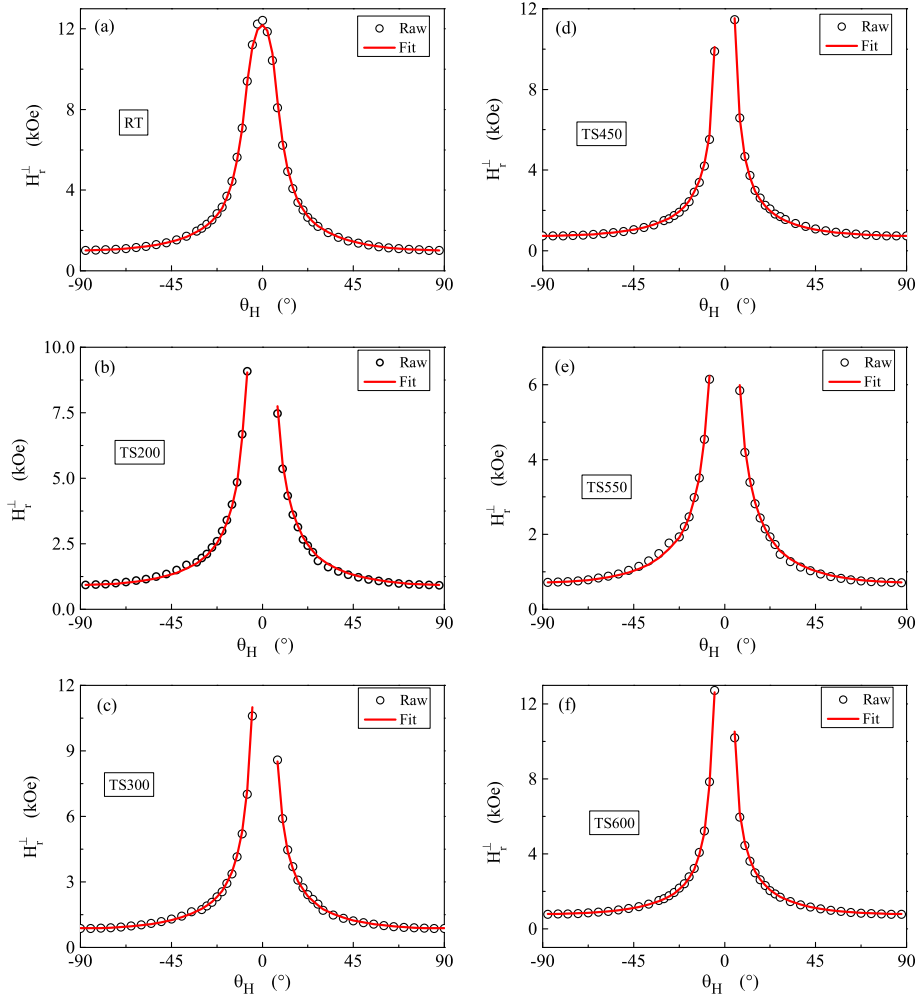}
\caption{Variation of resonance field, $H_{r}^{\perp}$, as a function of the field angle, $\theta_{H}$, for 50 nm CFS thin films at room temperature in the OP configuration. The continuous curves through the data points represent the theoretical fits based on Eqs.(5) and (6).}
\label{figure 9}
\vspace{-0.5cm}
\end{figure*}

\begin{figure*}[htbp]
\centering
\includegraphics[scale=1.0, trim = 0 0 0 0, clip, width=\linewidth]{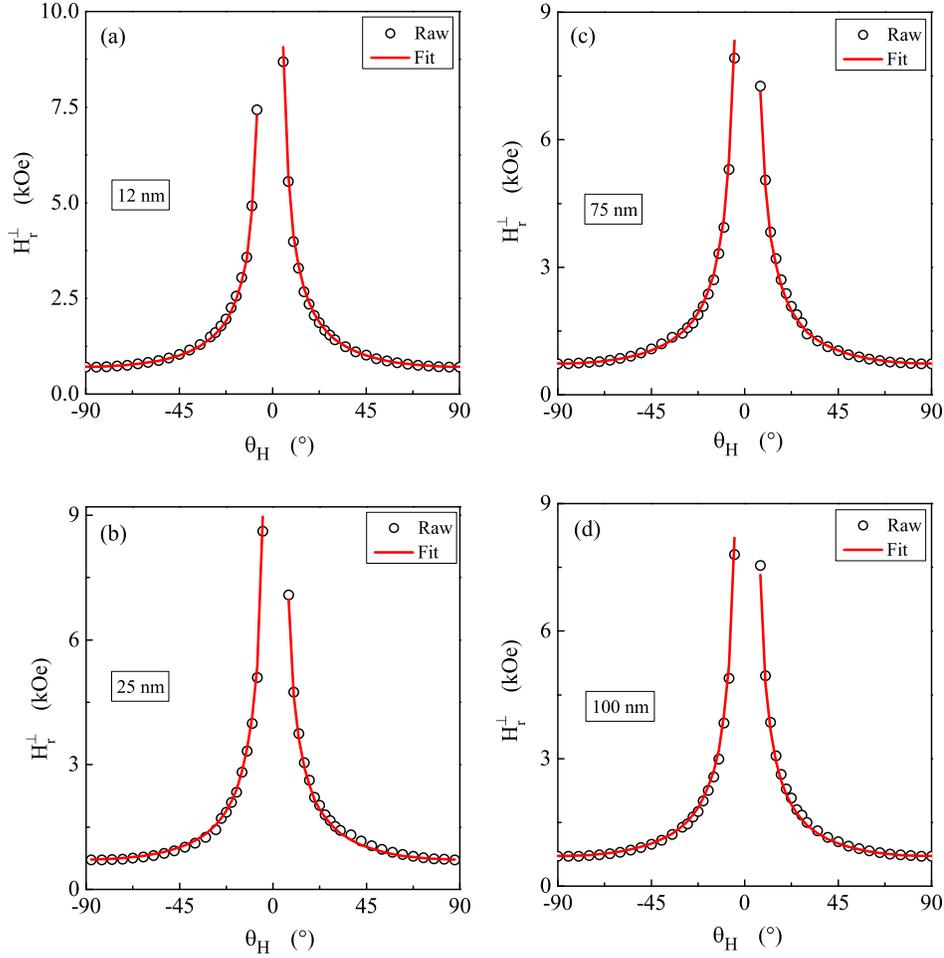}
\caption{The resonance field, $H_{r}^{\perp}$, as a function of $\theta_{H}$ for the CFS films with thickness in the range 12-100 nm at room temperature in the OP configuration. The continuous curves through the data points represent the theoretical fits based on Eqs.(5) and (6).}
\label{figure 10}
\vspace{-0.5cm}
\end{figure*}

\begin{figure*}[htbp]
\centering
\includegraphics[scale=1.0, trim = 0 250 20 170, clip, width=\linewidth]{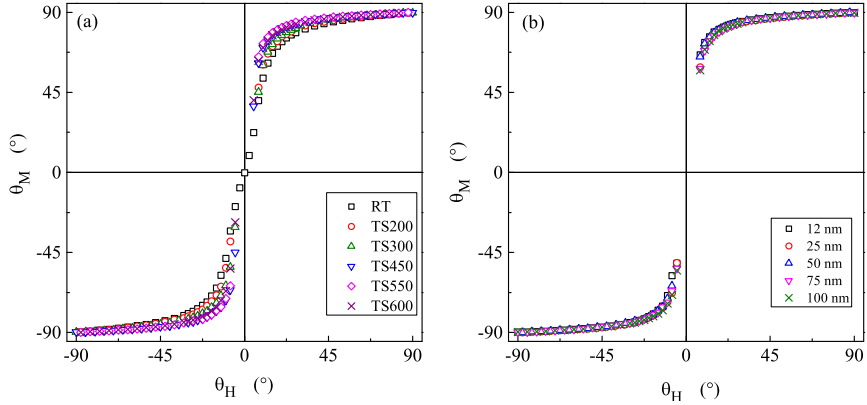}
\caption{Variation of the OP equilibrium magnetization angle, $\theta_{M}$, with the OP field angle, $\theta_{H}$, for (a) the 50 nm CFS thin films deposited at various T$_{S}$ and (b) the CFS films with thickness in the range 12-100 nm deposited at T$_{S}$ = 550 $^{\circ}$C.}
\label{figure 11}
\vspace{-0.5cm}
\end{figure*}

\begin{table}
\caption{\label{label} Values of the resonance field, H$_{r}^{\parallel}$, along the `in-plane' easy axis ($\varphi_{H}$ = 0$^{\circ}$) and hard axis ($\varphi_{H}$ = 90$^{\circ}$), magnetic anisotropy field, H$_{k}^{\parallel}$ = [H$_{r}^{\parallel}$($\varphi_{H}$ = 90$^{\circ}$) - H$_{r}^{\parallel}$($\varphi_{H}$ = 0$^{\circ}$)]/2, obtained from the fits, based on Eqs.(3) and (4), to the H$_{r}^{\parallel}(\varphi_{H})$ data for the 50 nm CFS films, deposited at different substrate temperatures. The values for M$_{s}$ and H$_{k}^{\parallel}$, obtained from the lineshape (LS) analysis, are also included for comparison.} 
\begin{indented}
\item[]\begin{tabular}{@{}llllll}
\br
Sample  & H$_{r}^{\parallel}$ &  H$_{r}^{\parallel}$  & H$_{k}^{\parallel}$ & M$_{s}$ & H$_{k}^{\parallel}$ \\
~~ & at $\varphi_{H}$ = 0$^{\circ}$ & at $\varphi_{H}$ = 90$^{\circ}$ & ~~  &  from LS fit  & from LS fit\\
~~ & (Oe) & (Oe) & (Oe) & (Gauss) & (Oe) \\
\mr
RT & 1038 & 1133 & 47.5 & 670 & 44 \\ 
TS200 & 956 & 1058 & 51.0 & 730 & 54 \\ 
TS300 & 904 & 965 & 30.5 & 801 & 32 \\ 
TS450 & 710 & 736 & 13.0 & 1074 & 13 \\  
TS550 & 711 & 727 & 8.0 & 1084 & 8 \\  
TS600 & 778 & 814 & 18.0 & 963 &  18 \\ 
\br
\end{tabular}
\end{indented}
\end{table}
\subsubsection{Angular variations of the resonance field and FMR linewidth}

The lowest resonance field (H$_{r}^{\parallel}$) for a particular $\varphi_{H}$ is identified as H$_{r}^{\parallel}$ corresponding to $\varphi_{H}$ = 0$^{\circ}$, which represents the `in-plane' easy axis of magnetization. In the `in-plane' (IP)~ configuration, the FMR spectra have been recorded at 5$^{\circ}$ intervals starting from $\varphi_{H}$ = 0$^{\circ}$. The variation of resonance field, H$_{r}^{\parallel}$, with $\varphi_{H}$, deduced from FMR spectra, are displayed as polar plots in figures 4 and 5 for the CFS films belonging to the series I and II, respectively. From these figures, it is evident that, for all the films, H$_{r}^{\parallel}(\varphi_{H})$ goes through minima at $\varphi_{H}$ = 0$^{\circ}$ and $\varphi_{H}$ = 180$^{\circ}$, signifying the `in-plane' easy axes of magnetization, and maxima at $\varphi_{H}$ = 90$^{\circ}$ and $\varphi_{H}$ = 270$^{\circ}$, denoting the `in-plane' hard axes of magnetization. H$_{r}^{\parallel}$ has the same minimum (maximum) values at $\varphi_{H}$ = 0$^{\circ}$ and 180$^{\circ}$ ($\varphi_{H}$ = 90$^{\circ}$ and 270$^{\circ}$). The lower value of H$_{r}^{\parallel}$ along the easy axes $\varphi_{H}$ = 0$^{\circ}$ and 180$^{\circ}$, and higher values along the hard axes $\varphi_{H}$ = 90$^{\circ}$ and 270$^{\circ}$ establishes the presence of {\it `in-plane' uniaxial anisotropy} in all the CFS films, irrespective of the strength of disorder and the film thickness. The observed uniaxial anisotropy may have its origin in the elongated grains, that form and grow due to the self-shadowing effect during the oblique-angle deposition of the CFS films \cite{HoshiJAP1996,McMichaelJAP2000,YeNano2002,FukumaJAP2009}. A self-consistent procedure \cite{Siruguri1996, BasheedJAP2011} is followed to fit the H$_{r}^{\parallel}(\varphi_{H})$ data, as illustrated below. In the first step, equilibrium `in-plane' magnetization angle ($\varphi_{M}$), corresponding to the field angle ($\varphi_{H}$), is extracted from the fit to H$_{r}^{\parallel}(\varphi_{H})$, based on Eq.(4), with the saturation magnetization (M$_{s}$) and anisotropy field (H$_{k}^{\parallel}$) fixed at the values obtained from the lineshape analysis. In the second step, Eq.(3) is used to fit H$_{r}^{\parallel}(\varphi_{H})$ with $\varphi_{M}$ fixed at the value, obtained in the first step, and by treating M$_{s}$ and H$_{k}^{\parallel}$ as the free fitting parameters. The M$_{s}$ and H$_{k}^{\parallel}$ values, obtained in second step, are used again in Eq.(4) to arrive at the new value of $\varphi_{M}$, which is then used in Eq.(3) to get the new M$_{s}$ and H$_{k}^{\parallel}$. This iterative procedure is repeated till the values of M$_{s}$, H$_{k}^{\parallel}$ and $\varphi_{M}$, for a given value of $\varphi_{H}$, do not change. The final fits to H$_{r}^{\parallel}(\varphi_{H})$, so obtained, are denoted as the continuous curves in Figs. 4 and 5. These theoretical fits represent the experimental data quite well for all the films and the fit parameters are listed in the Tables 1 and 2. The values of the anisotropy field calculated from the relation \cite{KaulJPCM1992, BabuPRB1992} H$_{k}^{\parallel}$ = [H$_{r}^{\parallel}(\varphi_{H} = 90^{\circ})$ - H$_{r}^{\parallel}(\varphi_{H} = 0^{\circ})$]/2, included in these tables for comparison, serve as a cross-check for the values of H$_{k}^{\parallel}$ obtained from the H$_{r}^{\parallel}(\varphi_{H})$ data. The reduction in H$_{k}^{\parallel}$ with increasing deposition temperature is directly related to the defect density in the multi-domain films \cite{VineetJAP2018}. The equilibrium magnetization angle, $\varphi_{M}$, is plotted against the corresponding field angle $\varphi_{H}$ for all the CFS films in figure 6. The observation that $\varphi_{M}$ $\cong$ $\varphi_{H}$ implies that the `in-plane' uniaxial anisotropy field is too small (clearly borne out by the values of H$_{k}^{\parallel}$ listed in the Tables 1 and 2) to counter the the external magnetic field ($\simeq$ H$_{r}^{\parallel}$) with the result that, irrespective of the value of $\varphi_{H}$, the magnetization vector points in the magnetic field direction.

The variation of linewidth, $\Delta H^{\parallel}$, for different CFS thin film samples, with $\varphi_{H}$ is shown in figures 7 and 8. Like the resonance field H$_{r}^{\parallel}$, $\Delta H^{\parallel}$, in all the cases, except for the films with $t$ = 12 nm, 25 nm and 50 nm deposited at the optimum substrate temperature T$_{S}$ = 550$^{\circ}$C (where $\Delta H^{\parallel}$ exhibits four-fold symmetry), shows two-fold symmetry. However, the values of $\varphi_{H}$, at which the minima and maxima in $\Delta H^{\parallel}(\varphi_{H})$ occur, do not match with those of $\varphi_{H}$ corresponding to the extrema in the resonance field, H$_{r}^{\parallel}$. The effect of disorder (T$_{S}$) gets clearly reflected in the $\Delta H^{\parallel}(\varphi_{H})$. The maximum variation of the `in-plane' linewidth [$\delta(\Delta H^{\parallel})$] is 10 Oe for the RT film, decreases systematically with T$_{S}$ and is the lowest ($\delta(\Delta H^{\parallel})$ $\sim$ 5 Oe) for the most ordered films of thickness in the range 12 nm - 100 nm. This proves that the RT film is more disordered compared to the completely ordered TS550 films of different thicknesses, as linewidth is a measure of magnetic inhomogeneity \cite{PlatowPRB1998, FarlePRB1996}. The fits to $\Delta H^{\parallel}(\varphi_{H})$, based on Eqs.(13), (15), (17) and (19), which consider the contributions to $\Delta H^{\parallel}$ due to LLG, two-magnon scattering (TMS), inhomogeneity in magnetization and angular spread of crystallite orientations \cite{BelmeguenaiPRB2013}, are denoted by solid curves in Figs. 7 and 8. These figures clearly demonstrate that, in the CFS thin films of varying thickness, a crossover from the four-fold symmetry (cubic anisotropy) to two-fold symmetry (uniaxial anisotropy) occurs as $t$ exceeds 50 nm. The two-fold (the term with prefactor $\Gamma_{2}$ in Eq.(15)) or four-fold symmetry (the term with prefactor $\Gamma_{4}$ in Eq.(15)) of $\Delta H^{\parallel}(\varphi_{H})$ mainly originates from the TMS, since LLG damping does not depend on $\varphi_{H}$ in the IP case, as is evident from Eqs.(13) and (15). 

At this stage, it should be emphasized that, since TMS is extremely sensitive to local structural distortions, defects and imperfections, the crossover from the four-fold to two-fold local symmetry, easily detected in $\Delta H^{TMS}$ (and so also in $\Delta H^{\parallel}(\varphi_{H})$), eludes detection in H$_{r}^{\parallel}(\varphi_{H})$ because the long-wavelength ($\textbf{q}$ = 0) uniform precession mode, excited at resonance, detects only the global two-fold symmetry (uniaxial anisotropy) but fails to recognize the presence of local structural inhomogeneities that lead to the four-fold to two-fold local symmetry crossover.

\begin{figure*}[htbp]
\centering
\includegraphics[scale=1.0, trim = 0 0 0 0, clip, width=\linewidth]{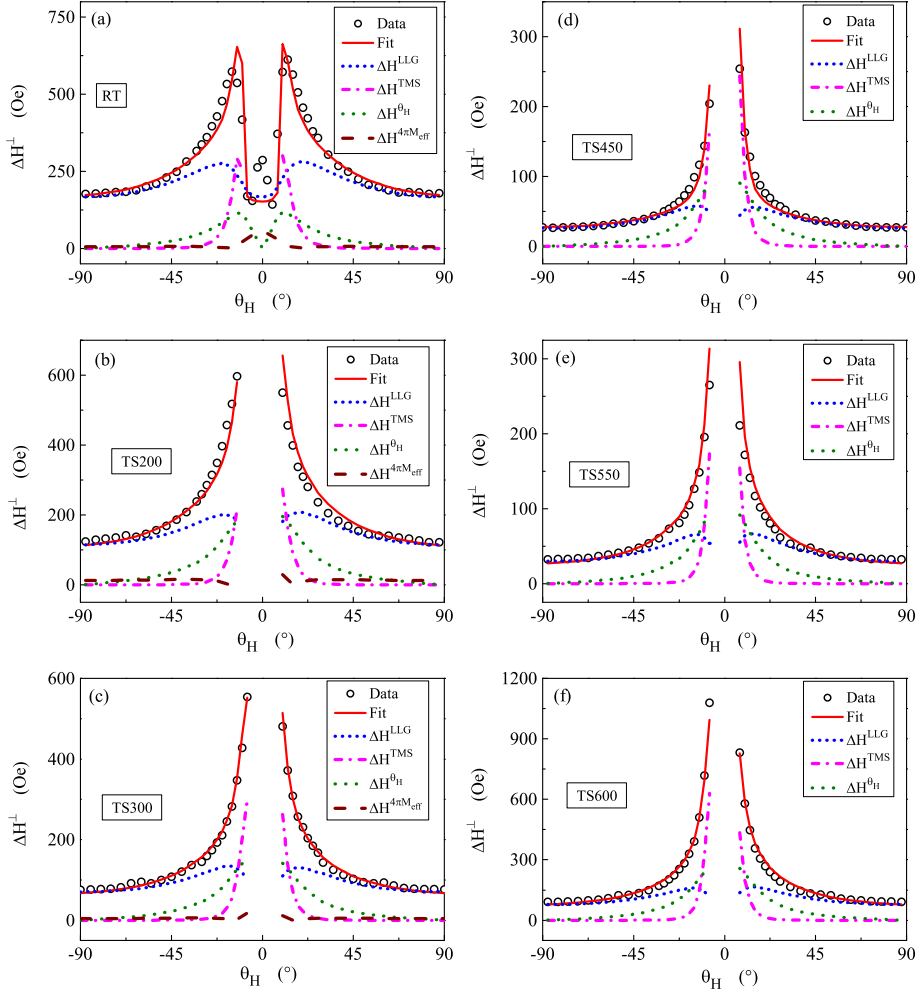}
\caption{ $\Delta H^{\perp}$ as a function of the field angle, $\theta_{H}$, for 50 nm CFS thin films at T = 300 K in the OP configuration. The continuous curves through the data points represent the theoretical fits based on Eqs.(14), (16), (18) and (20). The angular variation of the individual contributions, denoted by the dotted, dot-dashed and dashed lines, are also depicted.}
\label{figure 12}
\vspace{-0.5cm}
\end{figure*}

\begin{figure*}[htbp]
\centering
\includegraphics[scale=1.0, trim = 0 0 0 0, clip, width=\linewidth]{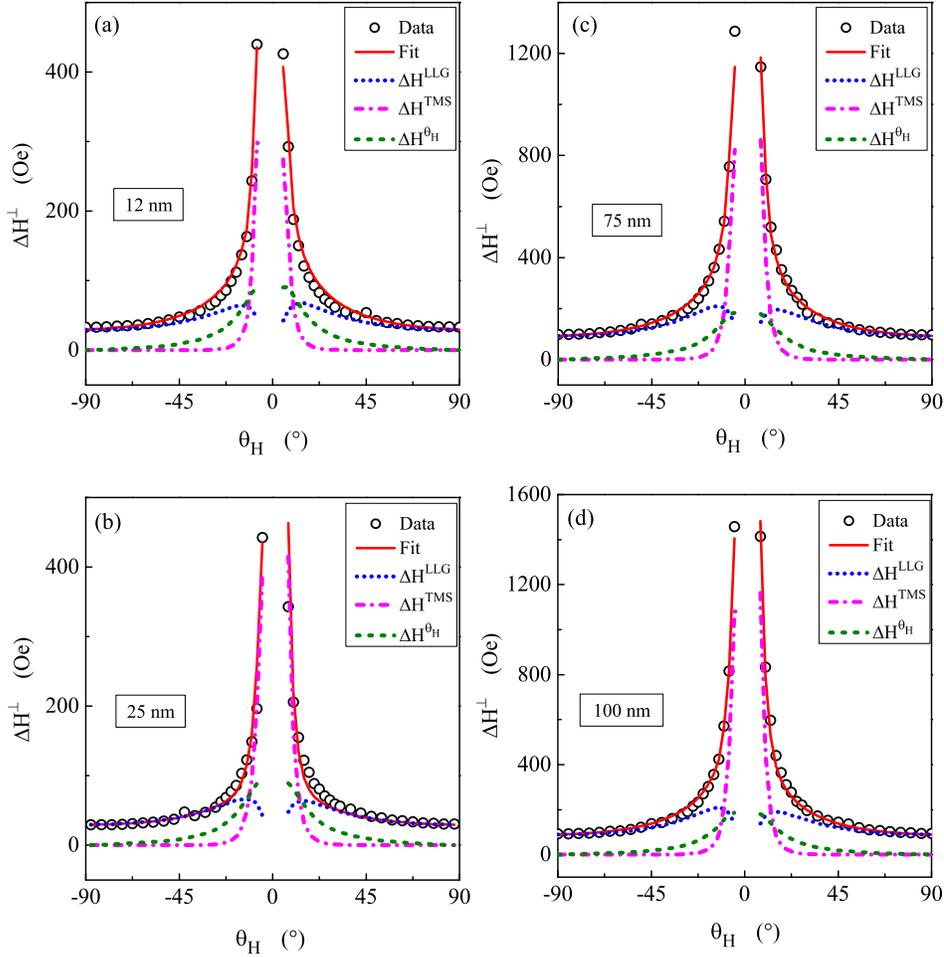}
\caption{$\Delta H^{\perp}$ as a function of $\theta_{H}$ for the CFS films of different thicknesses at T = 300 K in the OP configuration. The continuous curves through the data points represent the theoretical fits based on Eqs. (14), (16), (18) and (20). The angular variation of the individual contributions, denoted by the dotted, dot-dashed and dashed lines, are also depicted.}
\label{figure 13}
\vspace{-0.5cm}
\end{figure*}

\begin{figure*}[htbp]
\centering
\includegraphics[scale=1.0, trim = 0 100 0 100, clip, width=\linewidth]{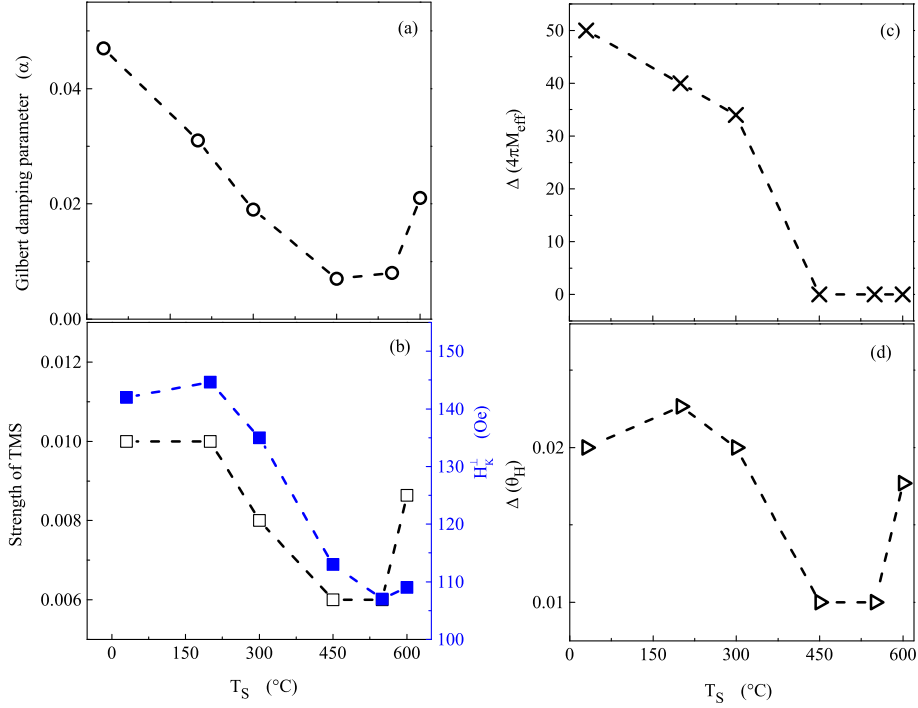}
\caption{(a) - (d) Variations of the Gilbert damping parameter ($\alpha$), strength of TMS, H$_{K}^{\perp}$, $\Delta(4 \pi M_{eff})$ and $\Delta(\theta_{H})$ with T$_{S}$, deduced from the $\Delta H^{\perp}$ versus $\theta_{H}$ fits, based on Eqs.(14), (16), (18) and (20).}
\label{figure 14}
\vspace{-0.5cm}
\end{figure*}

\begin{figure*}[htbp]
\centering
\includegraphics[scale=1.0, trim = 0 100 0 100, clip, width=\linewidth]{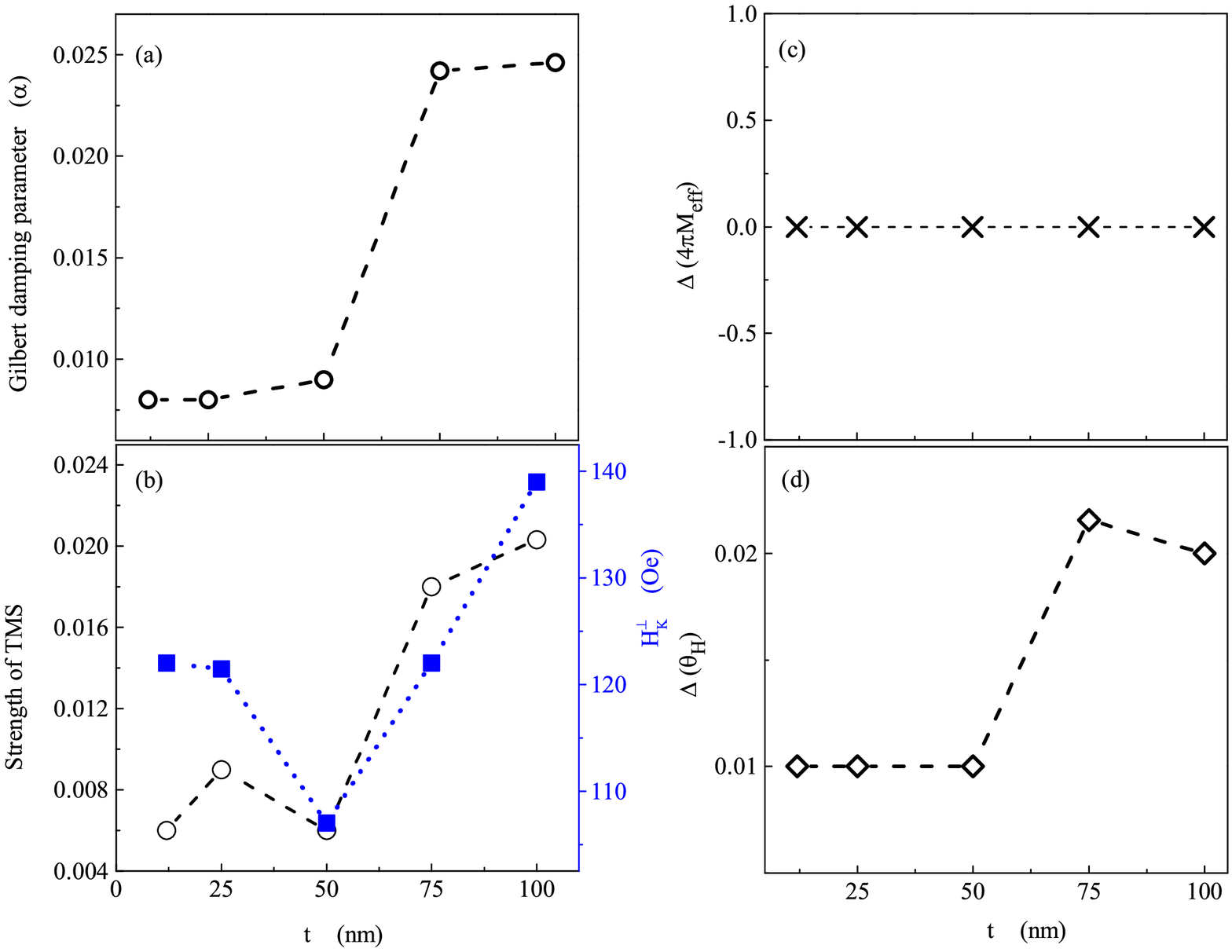}
\caption{(a) - (d) Variation of Gilbert damping parameter ($\alpha$), strength of TMS, H$_{K}^{\perp}$, $\Delta(4 \pi M_{eff})$ and $\Delta(\theta_{H})$ with thickness ($t$) deduced from the $\Delta H^{\perp}$ versus $\theta_{H}$ fit based on Eqs.(14), (16), (18) and (20).}
\label{figure 15}
\vspace{-0.5cm}
\end{figure*}

\begin{figure*}[htbp]
\centering
\includegraphics[scale=1.0, trim = 0 0 0 0, clip, width=\linewidth]{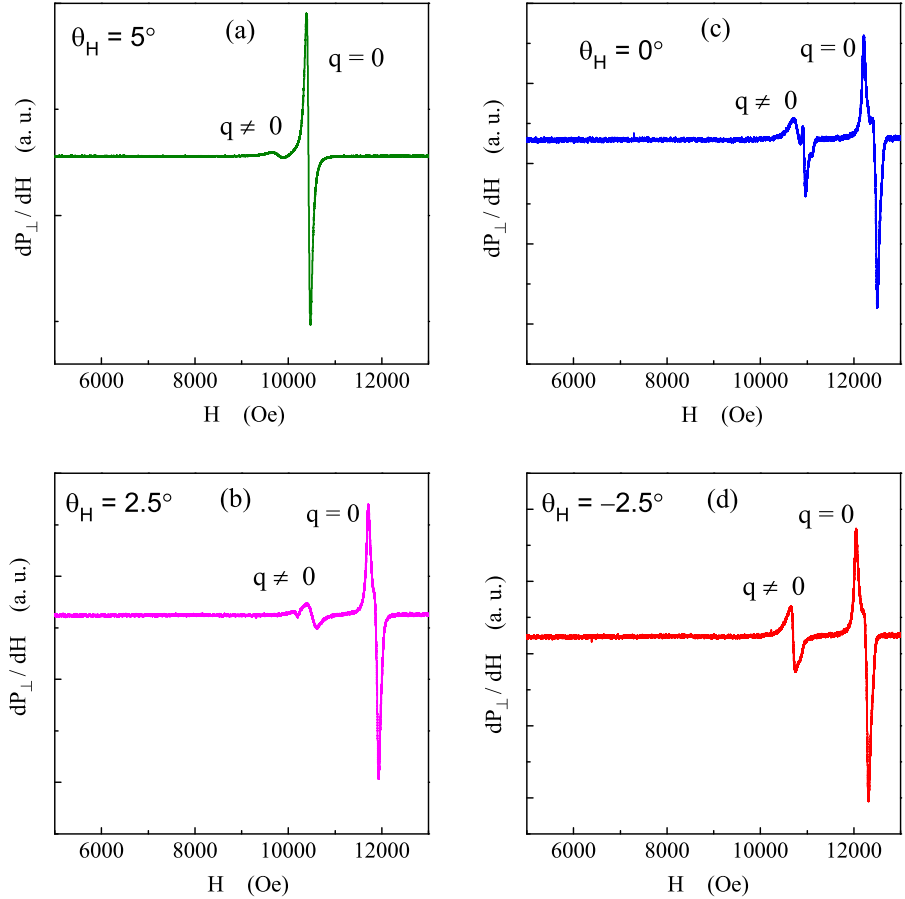}
\caption{uniform ($\textbf{q}$ = 0) and non-uniform ($\textbf{q}$ $\neq$ 0) precession spin-wave modes excited at resonance in the `out-of-plane' configuration in the RT CFS film.}
\label{figure 16}
\vspace{-0.5cm}
\end{figure*}
%
%
\begin{table}
\caption{\label{label} Values of the resonance field, H$_{r}^{\parallel}$, along the `in-plane' easy axis ($\varphi_{H}$ = 0$^{\circ}$) and hard axis ($\varphi_{H}$ = 90$^{\circ}$), magnetic anisotropy field, H$_{k}^{\parallel}$ = [H$_{r}^{\parallel}$($\varphi_{H}$ = 90$^{\circ}$) - H$_{r}^{\parallel}$($\varphi_{H}$ = 0$^{\circ}$)]/2, obtained from the fits, based on Eqs.(3) and (4), to the H$_{r}^{\parallel}(\varphi_{H})$ data for the CFS films of different thicknesses, deposited at 550$^{\circ}$C. The values for M$_{s}$ and H$_{k}^{\parallel}$, obtained from the lineshape (LS) analysis, are also included for comparison.}
\begin{indented}
\item[]\begin{tabular}{@{}llllll}
\br
t & H$_{r}^{\parallel}$ &  H$_{r}^{\parallel}$  & H$_{k}^{\parallel}$ & M$_{s}$ & H$_{k}^{\parallel}$ \\
~~ & at $\varphi_{H}$=0$^{\circ}$ & at $\varphi_{H}$=90$^{\circ}$ & ~~ &  from LS fit  & from LS fit\\
(nm)  & (Oe) & (Oe) & (Oe) & (Gauss) & (Oe) \\
\mr
12 & 704 & 728 & 12.0 & 1087 & 12 \\ 

25 & 709 & 727 & 9.0 & 1090 & 10 \\

50 &  711 & 727 & 8.0 & 1084 & 8 \\ 

75 &  731 & 754 & 11.5 & 1044 & 12 \\  
 
100 &  682 & 715 & 16.5 & 1120 & 16 \\  
\br
\end{tabular}
\end{indented}
\end{table}

\subsection{`Out-of-plane' (OP) configuration}

In the `out-of-plane' (OP) configuration, FMR spectra have been recorded at 5$^{\circ}$ ($\sim$2$^{\circ}$) intervals starting from $\theta_{H} = -90^{\circ}$ to +90$^{\circ}$ (near 0$^{\circ}$). The variations of the resonance field, H$_{r}^{\perp}$, with $\theta_{H}$ for the series I and II CFS thin films, deduced from the FMR spectra, are displayed in figures 9 and 10, respectively. H$_{r}^{\perp}$ has the same values at the angles $\theta_{H} = -90^{\circ}$ and $\theta_{H} = 90^{\circ}$, and increases rapidly so as to reach a value as high as $\simeq$ 13 kOe [the upper instrumental limit for the static field $\sim$ 13 kOe] as $\theta_{H} = 0^{\circ}$ is approached either from below or above. It follows from this angular variation of H$_{r}^{\perp}$ that the magnetization prefers to lie within the film plane for all the CFS films. When $\theta_{H} = -$90$^{\circ}$ or $\theta_{H} = 90^{\circ}$, the magnetic field (\textbf{H}) lies within the film plane and points along the easy axis of magnetization while \textbf{H} points along the film normal when $\theta_{H} = 0^{\circ}$. The resonance field at $\theta_{H} = 0^{\circ}$ is $\sim$ 12 kOe for the RT film but shifts to higher fields for the remaining films. A self-consistent procedure \cite{Siruguri1996, BasheedJAP2011} is followed to fit the H$_{r}^{\perp}(\theta_{H})$, based on Eq.(5) and Eq.(6), with saturation magnetization and anisotropy field as the free fitting parameters, in a way similar to that described earlier in the IP case. The above procedure yields the theoretical fits to the H$_{r}^{\perp}(\theta_{H})$, shown in Figs. 9 and 10 as continuous curves, that represent the experimental data (open circles) quite well for all the films. The equilibrium magnetization angle, $\theta_{M}$, is plotted against $\theta_{H}$, in figure 11. $\theta_{M}$ = $\theta_{H}$ at $\theta_{H} = -90^{\circ}$ or $= +90^{\circ}$, i.e., the magnetization vector points in the field direction only when $H$ lies within the film plane because the `in-plane' anisotropy field is too small compared to $H$ (as in the `in-plane' case). From $\theta_{H}$ = $\pm$ $90^{\circ}$ to $\theta_{H}$ $\simeq$ $\pm$ $20^{\circ}$, $\theta_{M}$ does not deviate significantly from $\pm$ $90^{\circ}$ and only when $\theta_{H}$ $\rightarrow$ 0$^{\circ}$, $\theta_{M}$ also approaches 0$^{\circ}$. This observation implies that, except for $\theta_{H}$ in the range $-20^{\circ}$ $\lesssim$ $\theta_{H}$ $\lesssim$ $+20^{\circ}$, the applied static magnetic field (= H$_{r}^{\perp}$) is not strong enough (compared to the `in-plane' anisotropy field) to extract the magnetization vector out of the film plane.

The variations of the OP-FMR linewidth, $\Delta H^{\perp}$, with $\theta_{H}$ for the series I and II CFS films are displayed in figures 12 and 13. For the RT film, $\Delta H^{\perp}$ initially increases up to $\theta_{H} = 10^{\circ}$ or $\theta_{H} = -10^{\circ}$ and then gives rise to a dip at $\theta_{H} = 0^{\circ}$. Such a dip at $\theta_{H} = 0^{\circ}$ is not observed for the remaining films because H$_{r}^{\perp}$ shifts to fields higher than the upper instrumental limit of $\sim$ 13 kOe. The Gilbert damping parameter can be extracted from the $\Delta H^{\perp}(\theta_{H})$ data as follows. In permalloy \cite{MizukamiJJAP2001} and Heusler alloy \cite{OoganeAPL2010, YangAPL2013, MikihikoJAP2007} thin films, it is reported that the LLG damping along with inhomogeneous broadening due to the distribution in local magnetization and the angular spread in crystallite misorientation, contribute to $\Delta H^{\perp}$. In the present study, however, we demonstrate that the observed $\Delta H^{\perp}(\theta_{H})$ in the CFS thin films cannot be reproduced unless the two-magnon scattering (TMS) is taken into consideration together with intrinsic and other extrinsic contributions.

The theory of TMS was originally proposed for the FMR line-broadening in magnetic insulators \cite{CochranPRB1989}. In many ferromagnetic metals, LLG damping alone could not account for the observed linewidth. Thus, the theory of TMS was extended to metallic ferromagnets by Heinrich et al. \cite{CochranPRB1989, HeinrichJAP1985}. The alternating microwave field excites magnon modes of different wave vectors ($\textbf{q} \neq 0$) that interact with one another and give rise to the TMS in metallic films. Another possible mechanism of line-broadening is the exchange conductivity contribution, which is ruled out as the skin depth ($\sim$ 1~$\mu$m) is larger than the thickness of the CFS films \cite{CochranPRB1989}. In the present study, the angular variation of $\Delta H^{\perp}$ is fitted considering the LLG contribution ($\Delta H^{LLG}$), the line-broadening due to inhomogeneity in magnetization ($\Delta H^{4 \pi M_{eff}}$), line-broadening arising from the angular spread in the crystallite misorientation ($\Delta H^{\theta_{H}}$) and TMS ($\Delta H^{TMS}$). The equilibrium `out-of-plane' magnetization angle, obtained from the fit to H$_{r}^{\perp}(\theta_{H})$, is used to calculate the above-mentioned contributions to linewidth. The theoretical fits to $\Delta H^{\perp}(\theta_{H})$, based on the Eqs.(14), (16), (18) and (20), are depicted by the solid curves in figures 12 and 13. The individual linewidth contributions are also plotted in these figures. The fit parameters such as the Gilbert damping parameter ($\alpha$), the magnitude of TMS, H$_{K}^{\perp}$, $\Delta(4 \pi M_{eff})$ and $\Delta \theta_{H}$ are plotted as functions of T$_{S}$ and $t$ in the panels (a) - (d) of the figures 14 and 15.

The following inferences can be drawn from the theoretical fits to $\Delta H^{\perp}(\theta_{H})$, shown in Figs.12 and 13.

(i) The LLG and $\Delta H^{\theta_{H}}$ contributions increase slowly as $\theta_{H} \rightarrow 30^{\circ}$ or $-30^{\circ}$ and then give rise to a dip around $\theta_{H} = 0^{\circ}$. Experimentally, $\Delta H^{\perp}$ shows a dip at $\theta_{H} = 0^{\circ}$ only in the RT film. This dip basically originates from the LLG and $\Delta H^{\theta_{H}}$ contributions. However, such a dip is not observed in the remaining films because of the shift in the resonance field to higher fields. While the two-magnon scattering is entirely responsible for the divergence in $\Delta H^{\perp}$ as $\theta_{H} \rightarrow 0^{\circ}$, the LLG mechanism largely determines $\Delta H^{\perp}$ in the $\theta_{H}$ intervals $-90^{\circ}$ to $-45^{\circ}$ and $+45^{\circ}$ to $+90^{\circ}$. With increasing T$_{S}$, the Gilbert (LLG) damping parameter, $\alpha$, decreases  from the value 0.047 for the RT film to 0.0078 for the TS450 and TS550 films (Fig.14(a)) whereas $\alpha$ jumps from 0.008 for the CFS film with $t$ = 50 nm to 0.024 for $t$ = 75 nm (Fig.15(a)). To understand the observed variations of $\alpha$ with T$_{S}$ and $t$, use is made of the torque correlation model \cite{Kambersky1970,SakumaJOPD2015}, which yields the relation 

\begin{eqnarray}
\fl
\alpha = (1/\gamma ~ M_{s} ~ \tau)~ \mu_{B}^{2} ~ N(E_{F})~ (g - 2)^{2} 
\label{Eq 21}  
\end{eqnarray}
where $N(E_{F})$ is the total spin ($S\uparrow$ and $S\downarrow$) density of states at the Fermi level $E_{F}$ and $\tau$ is the conduction electron scattering time. Considering that $\tau^{-1}$ is proportional to the electrical resistivity ($\rho$) and our finding that the factor $(g - 2)^{2}$ (which is a measure of the orbital magnetic moment) does not change significantly with disorder and/or film thickness, $t$, it follows from Eq.(21) that $\alpha$ $\sim$ ($\rho/M_{s}$) $N(E_{F})$. Use of the variations with T$_{S}$ (or disorder) and $t$ of $M_{s}$, observed in this work, and of $\rho$, reported previously in \cite{BKH2017,BKH2019}, in Eq.(21) permits us to conclude that the observed $\alpha(T_{S})$ and $\alpha(t)$ can be mimicked only when $N(E_{F})$ increases (decreases) with disorder (T$_{S}$) and exhibits an abrupt jump at $t~\gtrsim~50$nm. Consistent with our finding, the first principles calculations \cite{SakumaJOPD2015}, based on the torque correlation model \cite{Kambersky1970}, clearly bear out that    $N(E_{F})$ (and hence $\alpha$) increases with increasing disorder strength in the Co-based Heusler alloys $Co_{2}MnAl$, $Co_{2}MnSi$ and $Co_{2}FeSi$.

(ii) The extrinsic linewidth contribution due to inhomogeneity in magnetization, $\Delta(4 \pi M_{eff})$, shows up only for $\theta_{H}$ very close to $0^{\circ}$ and that too for the amorphous films alone (Fig.12(a) - (c)), decreases with increasing T$_{S}$, and vanishes for the crystalline films irrespective of their thickness (Fig.14(c) and Fig.15(c)). It immediately follows from this result that magnetization is more inhomogeneous in the amorphous films than in the crystalline counterparts, as expected. (iii) From Figs.12 and 13, it is evident that the TMS makes negligible contribution to $\Delta H^{\perp}(\theta_{H})$ in the ranges $- 90^{\circ} \leq \theta_{H} \lesssim  - 30^{\circ}$ and $30^{\circ} \lesssim \theta_{H} \leq 90^{\circ}$, but in the $\theta_{H}$ range $\gtrsim  - 30^{\circ}$ to $\lesssim  + 30^{\circ}$, TMS increases rapidly. The TMS contribution is larger in the amorphous films than in the crystalline films because, unlike crystalline ferromagnets, amorphous counterparts have a spatial distribution in the local exchange interaction, local magnetization and local anisotropy fields: all such local inhomogeneities are additional sources of two-magnon scattering \cite{HeinrichJAP1985}. As $\theta_{H} \rightarrow 0^{\circ}$, TMS scattering is more effective since the spin wave modes with wave vector $\textbf{q} \neq 0$ are excited in the perpendicular configuration ($\theta_{H} = 0^{\circ}$) rather than in the `in-plane' ($\theta_{H} = 90^{\circ}$) configuration for a fixed microwave frequency. We could clearly observe such resonant microwave field-excited spin-wave modes with wave vector $\textbf{q} \neq 0$, apart from the uniform precession spin-wave modes with wave vector $\textbf{q} = 0$, in the FMR spectra taken on the RT Co$_{2}$FeSi (CFS) Heusler alloy 50 nm thick films in the `out-of-plane' sample configuration, as is evident from figure 16. This is so because, in this sample alone, the resonance corresponding to the $\textbf{q} = 0$ uniform precession mode could be detected right up to $\theta_{H} = 0^{\circ}$.     

Lastly, we comment on the effect of spin-pumping on $\alpha$. For uniform `out-of-plane' precession, excited in a ferromagnetic (FM) layer at the resonant microwave frequencies in a FMR experiment, the spins, pumped from a FM layer into a non-ferromagnetic (NM) layer, diffuse into the NM layer in a direction perpendicular to the FM/NM interface. For each FM/NM interface, the spin-pumping increases $\alpha$ from its (spin-pumping free) FM bulk value, $\alpha_{0}$, by the amount \cite{TserkovnyakPRB2002, PolianskiPRL2004, BooneJAP2013}

\begin{eqnarray}
\fl
\Delta\alpha = \alpha - \alpha_{0} = \frac{g~\mu_{B}}{4\pi~M_{s}~t_{FM}} ~ g^{\uparrow \downarrow}
\label{Eq 22}  
\end{eqnarray}

where $t_{FM}$ is the thickness of the FM layer and $g^{\uparrow \downarrow}$ is the \textit{effective} spin-mixing conductance. In the present case, the relevant FM/NM interface is between the Co$_{2}$FeSi thin film and the 2 nm thick Ta cap layer. Our finding that $\alpha$ has the same value (within the uncertainty limits) for the 50 nm thick CFS films, deposited at RT, with and without Ta cap layer, rules out a possible spin-pumping contribution to $\alpha$. This conclusion is further supported by the observation that, for practically constant (film thickness-independent) values of $M_{s}$ and $g^{\uparrow \downarrow}$, Eq.(22) predicts that the spin-pumping contribution to $\alpha$ decreases with increasing $t_{FM}$ whereas a completely opposite trend is observed (Fig.15(a)). 

\section{SUMMARY AND CONCLUSIONS}

FMR spectra have been taken at different static magnetic field angles in the `in-plane' and `out-of-plane' thin film configurations at room temperature on 50 nm thick Co$_{2}$FeSi (CFS) Heusler alloy thin films, deposited at the Si(111) substrate temperatures (T$_{S}$) ranging from room temperature to 600$^\circ$C, and on the CFS films with thickness ($t$) in the range 12 nm to 100 nm, deposited at the optimum substrate temperature T$_{S}$ = 550$^\circ$C. Use is made of an elaborate data analysis of the angular variations of the resonance field ($H_{r}$) and linewidth ($\Delta H$) in the `in-plane' (IP, $\parallel$) and `out-of-plane' (OP, $\perp$) configurations, that, besides the Landau-Lifshitz-Gilbert damping (LLG, $\Delta H^{LLG}$), considers all possible extrinsic FMR linewidth contributions such as the two-magnon scattering (TMS, $\Delta H^{TMS}$), the line-broadening due to inhomogeneity in magnetization ($\Delta H^{4 \pi M_{eff}}$), and that ($\Delta H^{\theta_{H}}$) arising from the angular spread in the crystallite misorientation. The results of this analysis enable us to draw the following conclusions unambiguously.

(I) Regardless of the disorder strength and CFS film thickness, \textit{global} IP uniaxial anisotropy exists whose strength decreases with improving crystalline order in the films. (II) The OP uniaxial anisotropy field, $H_{k}^{\perp}$, is an order of magnitude higher than the IP counterpart, $H_{k}^{\parallel}$. (III) In the CFS thin films of varying thickness, a crossover from the `in-plane' \textit{local} four-fold symmetry (cubic anisotropy) to \textit{local} two-fold symmetry (uniaxial anisotropy) occurs as $t$ exceeds 50 nm. (IV) In both IP and OP cases, Landau$-$Lifshitz$-$Gilbert damping and two-magnon scattering make dominant contributions to the FMR linewidth. (V) The two-magnon scattering has larger magnitude in the amorphous films than in the crystalline ones. (VI) Gilbert damping parameter, $\alpha$, decreases monotonously from 0.047 to 0.0078 with decreasing disorder strength (increasing T$_{S}$) and jumps from 0.008 for the CFS film with $t$ = 50 nm to 0.024 for the film with $t$ = 75 nm. Such variations of $\alpha$ with T$_{S}$ and $t$ are understood in terms of the changes in the total (spin-up and spin-down) density of states at the Fermi level caused by the disorder and film thickness. (VII) Spin pumping across the Co$_{2}$FeSi film/Ta cap-layer interface makes negligible contribution to $\alpha$. (VIII) Our results suggest that disorder and/or the film thickness can be used as control parameters to tune $\alpha$ in the CFS thin films to make them suitable for a given spintronics application.

\section*{ACKNOWLEDGMENTS}
B. K. Hazra acknowledges the financial assistance (SRF) from UGC-BSR and thanks the Central Instruments Laboratory, University of Hyderabad, for permitting the use of the JEOL-FA200 Electron Spin Resonance Spectrometer. This work was supported by Indian National Science Academy under Grant No.: SP/SS/2013/1971.

\end{document}